\documentclass{wscpaperproc}
\usepackage{latexsym}
\usepackage{graphicx}
\usepackage{mathptmx}

\usepackage{amsmath}
\usepackage{amsfonts}
\usepackage{amssymb}
\usepackage{amsbsy}
\usepackage{amsthm}

\usepackage{subfigure}
\usepackage{lipsum,graphicx,multicol}

\usepackage[pdftex,colorlinks=true,urlcolor=blue,citecolor=black,anchorcolor=black,linkcolor=black]{hyperref}

\newtheoremstyle{wsc}
{3pt}
{3pt}
{}
{}
{\bf}
{}
{.5em}
{}

\theoremstyle{wsc}
\newtheorem{theorem}{Theorem}

\newtheorem{assumption}{Assumption}

\begin{document}

%
%

\pagestyle{fancyplain}

\thispagestyle{plain}
\firstPageHead{}

\chead{\fancyplain{}{\itshape Huang, Chen and Zhu}}

\rhead{}
\cfoot{}
\renewcommand{\headrulewidth}{0pt} 

\makeatletter
\let\@internalcite\cite
\def\cite{\def\@citeseppen{-1000}%
    \def\@cite##1##2{(##1\if@tempswa , ##2\fi)}%
    \def\citeauthoryear##1##2##3{##1 ##3}\@internalcite}
\def\citeNP{\def\@citeseppen{-1000}%
    \def\@cite##1##2{##1\if@tempswa , ##2\fi}%
    \def\citeauthoryear##1##2##3{##1 ##3}\@internalcite}
\def\citeN{\def\@citeseppen{-1000}%
    \def\@cite##1##2{##1\if@tempswa, ##2)\else{}\fi}%
    \def\citeauthoryear##1##2##3{##1 (##3)}\@citedata}
\def\citeA{\def\@citeseppen{-1000}%
    \def\@cite##1##2{(##1\if@tempswa , ##2\fi)}%
    \def\citeauthoryear##1##2##3{##1}\@internalcite}
\def\citeANP{\def\@citeseppen{-1000}%
    \def\@cite##1##2{##1\if@tempswa , ##2\fi}%
    \def\citeauthoryear##1##2##3{##1}\@internalcite}
\def\shortcite{\def\@citeseppen{-1000}%
    \def\@cite##1##2{(##1\if@tempswa , ##2\fi)}%
    \def\citeauthoryear##1##2##3{##2 ##3}\@internalcite}
\def\shortciteNP{\def\@citeseppen{-1000}%
    \def\@cite##1##2{##1\if@tempswa , ##2\fi}%
    \def\citeauthoryear##1##2##3{##2 ##3}\@internalcite}
\def\shortciteN{\def\@citeseppen{-1000}%
    \def\@cite##1##2{##1\if@tempswa, ##2\else{}\fi}%
    \def\citeauthoryear##1##2##3{##2 (##3)}\@citedata}
\def\shortciteA{\def\@citeseppen{-1000}%
    \def\@cite##1##2{(##1\if@tempswa , ##2\fi)}%
    \def\citeauthoryear##1##2##3{##2}\@internalcite}
\def\shortciteANP{\def\@citeseppen{-1000}%
    \def\@cite##1##2{##1\if@tempswa , ##2\fi}%
    \def\citeauthoryear##1##2##3{##2}\@internalcite}
\def\citeyear{\def\@citeseppen{-1000}%
    \def\@cite##1##2{(##1\if@tempswa , ##2\fi)}%
    \def\citeauthoryear##1##2##3{##3}\@citedata}
\def\citeyearNP{\def\@citeseppen{-1000}%
    \def\@cite##1##2{##1\if@tempswa , ##2\fi}%
    \def\citeauthoryear##1##2##3{##3}\@citedata}
%
%
%
\def\@citedata{%
    \@ifnextchar [{\@tempswatrue\@citedatax}%
                  {\@tempswafalse\@citedatax[]}%
}

\def\@citedatax[#1]#2{%
\if@filesw\immediate\write\@auxout{\string\citation{#2}}\fi%
  \def\@citea{}\@cite{\@for\@citeb:=#2\do%
    {\@citea\def\@citea{, }\@ifundefined
       {b@\@citeb}{{\bf ?}%
       \@warning{Citation `\@citeb' on page \thepage \space undefined}}%
{\csname b@\@citeb\endcsname}}}{#1}}%

%
\def\@citex[#1]#2{%
\if@filesw\immediate\write\@auxout{\string\citation{#2}}\fi%
  \def\@citea{}\@cite{\@for\@citeb:=#2\do%
    {\@citea\def\@citea{, }\@ifundefined
       {b@\@citeb}{{\bf ?}%
       \@warning{Citation `\@citeb' on page \thepage \space undefined}}%
{\csname b@\@citeb\endcsname}}}{#1}}%

%
\def\@biblabel#1{}
\makeatother



\newdimen\bibindent
\bibindent=0.0em
\def\thebibliography#1{\section*{\refname}\list
   {}{\settowidth\labelwidth{[#1]}
   \leftmargin\parindent
   \itemindent -\parindent
   \listparindent \itemindent
   \itemsep 0pt
   \parsep 0pt}
   \def\newblock{}
   \sloppy
   \sfcode`\.=1000\relax}


\setlength{\baselineskip}{12.7pt}

\title{\uppercase{Distributed and Optimal Resilient Planning of Large-Scale Interdependent Critical Infrastructures}}

\author{
Linan Huang,
Juntao Chen,
Quanyan Zhu\\
\vspace{12pt}\\
Department of Electrical and Computer Engineering\\
 New York University\\
 2 MetroTech Center, Brooklyn, NY, 11201, USA
}

\maketitle

\section*{ABSTRACT}
The complex interconnections between heterogeneous critical infrastructure sectors make the system of systems (SoS) vulnerable to natural or human-made disasters and lead to cascading failures both within and across sectors. 
Hence, the robustness and resilience of the interdependent critical infrastructures (ICIs) against extreme events are essential for delivering reliable and efficient services to our society. 
To this end, we first establish a holistic probabilistic network model to model the interdependencies between infrastructure components. To capture the underlying failure and recovery dynamics of ICIs, we further propose a Markov decision processes (MDP) model in which the repair policy determines a long-term performance of the ICIs. To address the challenges that arise from the curse of dimensionality of the MDP, we reformulate the problem as an approximate linear program and then simplify it using factored graphs. 
We further obtain the distributed optimal control for ICIs under mild assumptions. Finally, we use a case study of the interdependent power and subway systems to corroborate the results and show that the optimal resilience resource planning and allocation can reduce the failure probability and mitigate the impact of failures caused by natural or artificial disasters.

\section{INTRODUCTION}

Presidential Policy Directive 21 (PPD-21) identifies 16 critical infrastructure sectors including energy, communication and transportation systems as so vital that their malfunctions and incapacitation can lead to an enormous economic loss and public safety threat  \cite{obama2013presidential}. 
Driven by the recent advances in information and communication technologies (ICTs) and the Internet of Things (IoTs), these sectors become highly interconnected, enabling faster information exchange and a higher level of situational awareness for real-time operations.
For example, \shortcite{zimmerman2016promoting} and \shortcite{zimmerman2018network} investigate the interdependency between food, energy and water. \shortcite{nwu020} shows that on the one hand, supervisory control and data acquisition (SCADA) systems control the power generation, transmission, and distribution according to the monitoring information from the communication sector. On the other hand, the energy sector provides the power to guarantee the normal operation of the SCADA systems. 

\begin{figure*}[]
\begin{multicols}{2}
\includegraphics[height=0.5\linewidth, width=\linewidth]{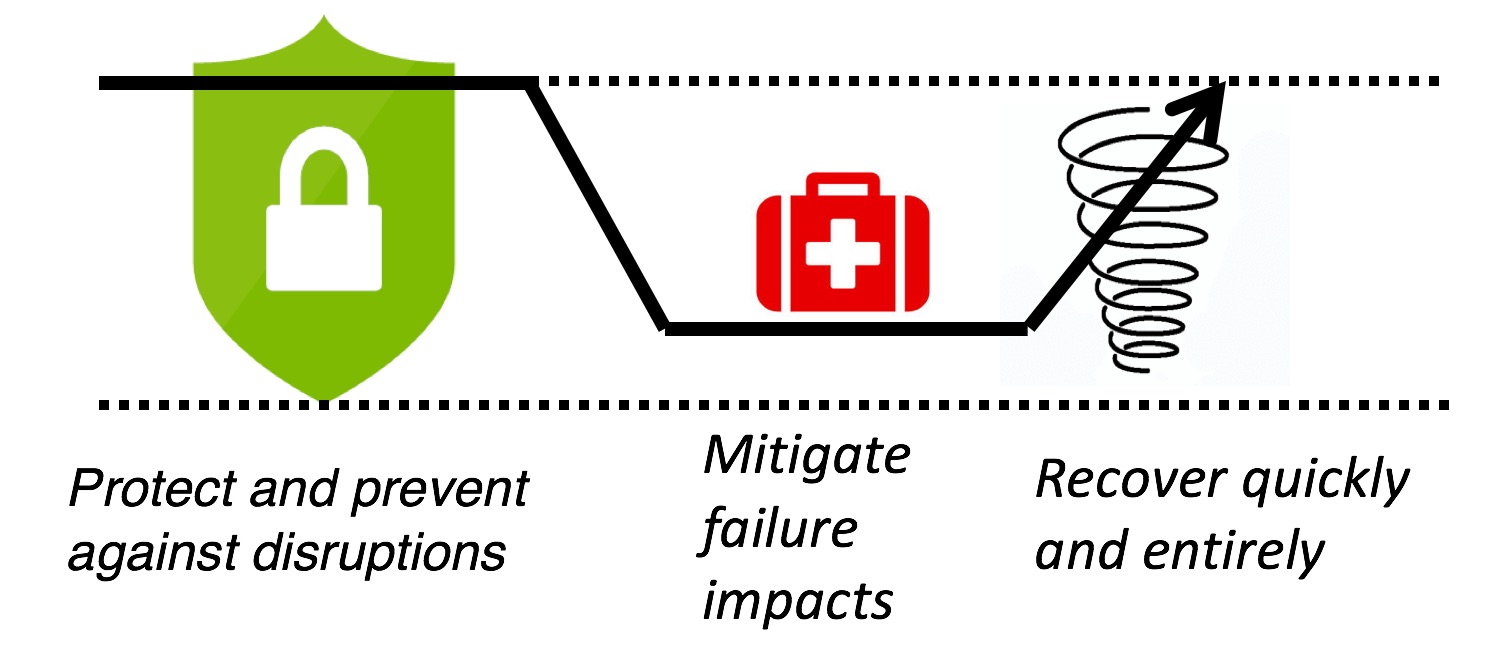}\par\caption{The systematic resilience consists of three phases. First is the robustness against disruptions to prevent failures. Second is the impact mitigation to guarantee the essential services under failures. Third is the prompt response and recovery. A paradigm shift from protections to the mitigation and recovery is indispensable for catastrophes.}
\label{Resilience}

    \includegraphics[height=0.5\linewidth,  width=\linewidth]{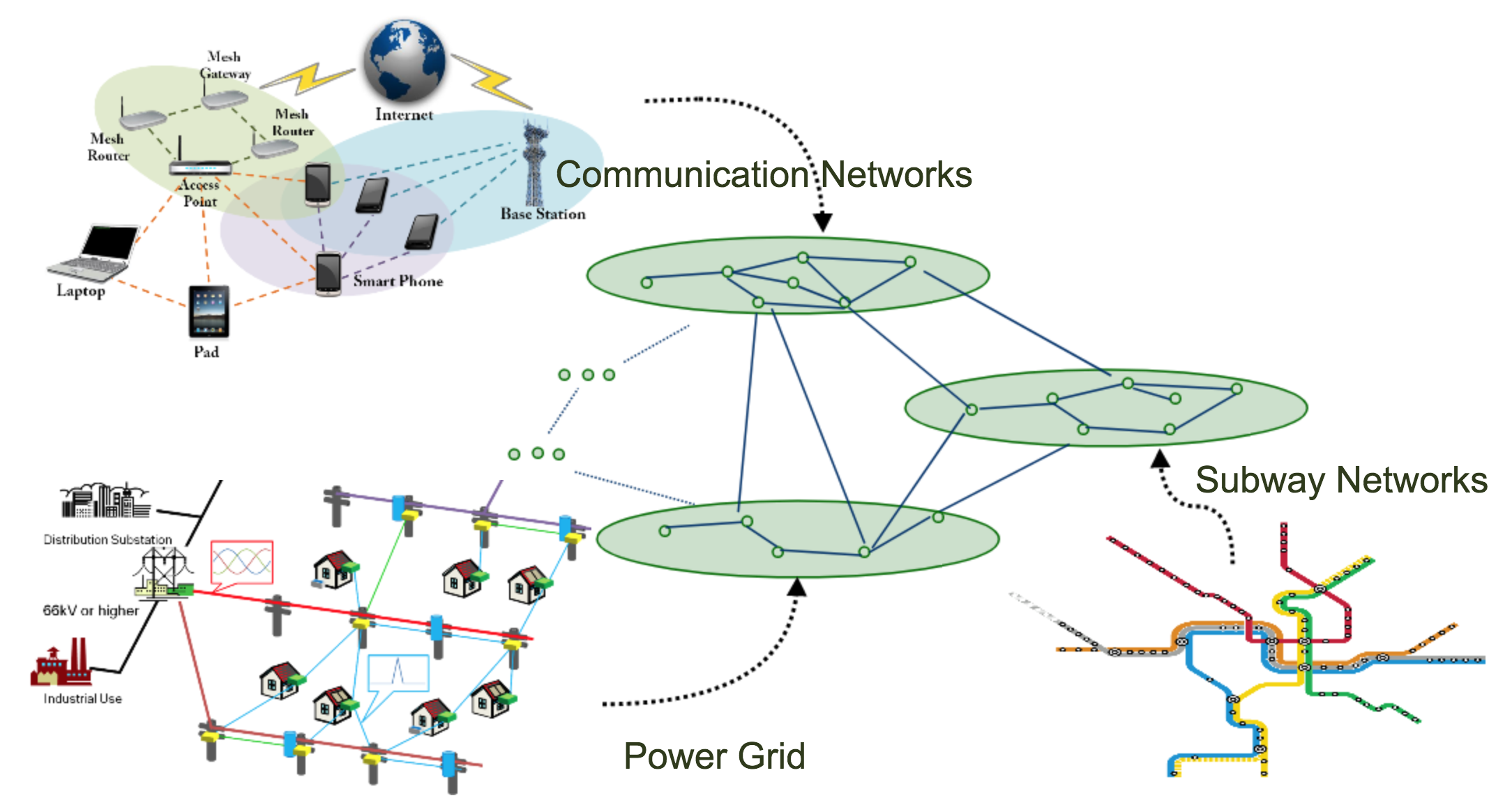}\par\caption{Multi-layer networks representing interdependent infrastructures such as power grid, subway and communication networks. They can be viewed as a large-scale aggregated network. The directed link between two nodes indicates either physical, cyber or logical influences from the source to the sink.}
    \label{multilayer}
\end{multicols}
\end{figure*}

Protection of interdependent critical infrastructures (ICIs) against natural or human-made disasters is essential for providing support to a reliable and sustainable economy. 
The increasing interdependencies, however, make systems vulnerable to these catastrophes because a failure of one system will propagate to others and cause cascading failures without appropriate, rapid responses. 
Lessons of the September 11 attack, hurricanes Sandy and  Irma have highlighted that protection and prevention against such disruptions are not always possible. Hence a paradigm shift to emphasize the preparedness and response is indispensable to enhance the resilience of ICIs as shown in Fig. \ref{Resilience}.
 
Many works have investigated the modeling, simulation, and design of interdependent critical infrastructures \shortcite{ouyang2014review,zimmerman2017conceptual}. 
Network flows \shortcite{lee2007restoration}  and  the interacting dynamic model \shortcite{rosato2008modelling} are used to assess and manage the risks.
Built upon the modeling of interdependent infrastructures,  \shortcite{yuan2015resilient} focuses on designing a resilient control for power grid. \shortcite{6603998} has proposed an L1Simplex architecture to address both physical and software failures in the cyber-physical systems. 
If an intelligent attacker rather than a random accident causes the network component failure, \shortcite{zhu2015game,huang2017large} apply dynamic game frameworks to incorporate the strategic participants who adapt to the defense policy.

Despite these pioneering works, challenges still remain to build a holistic model that captures not only distinct characteristics of individual infrastructures but also complex dependencies among sectors due to geographical, logical, cyber-physical, or human connections \shortcite{rinaldi2001identifying}. 
Without a holistic framework, we cannot integrate the data or models of individual sectors to make a globally optimal planning and defense decisions. For example,  system operators should not only make plans for normal operations of their own system but also be concerned about the possible failures of other dependent sectors and prepare for the potential influences. 
The framework should also be general enough to include various sectors and straightforward enough to enable real-time plannings for large-scale SoS, which motivates the MDP model in Section \ref{model} to abstract the complex networks, unify the multi-layer interdependences, and design resilient control mechanisms. 

Moreover, the \textbf{probabilistic} and \textbf{persistent} transition of system status requires system operators to prepare for the uncertainties in a dynamic perspective and avoid the following two extremes. 
First is the blind optimism that satisfies with the current normal operation and makes no sufficient contingency plan for probabilistic failures. 
Second is the pursuit of absolute security that would result in a huge inspection and maintenance cost. 
Thus, it is crucial to make \textbf{proactive} and \textbf{cost-effective} controls based on the real-time status of SoS to reduce the long-term risk economically and globally.

In this paper, we first establish a holistic network model to capture the relationships between nodes within an individual infrastructure and across the infrastructures. We use a probabilistic model to assess the impact of the couplings between different components in a system,  in which the failure probability of one component depends not only on its state but also on state of its connected ones.
One fundamental characteristic of the infrastructures is their dynamics in real-time operations \shortcite{zimmerman2018network}. For example, the load on a bus in a power grid changes due to the time-varying demands, and the bus would encounter a load shedding if the demand exceeds the supply. The concept of resilience involves the phase of recovery of  infrastructure after disastrous events. This process is naturally a dynamic one \shortcite{ChenSigmetrics2017}.  Hence, building on the probabilistic failure model, we propose a Markov decision processes (MDP) model to take into account the underlying dynamic processes of ICIs and develop resilient mechanisms for heterogeneous interdependent infrastructures. 

The MDP model has naturally a large state-space due to the large-scale feature of infrastructure networks and exponential increases of the system states   with respect to the number of components \shortcite{huang2017large,hayel2015resilient}. The curse of dimensionality of the MDP makes the control design computationally intractable even in a moderate-size system with 15 nodes. 
To this end, we reduce the dimension by first using basis functions to reformulate the problem as an approximate linear program (ALP) and then leveraging the sparsity of the interconnections to simplify the ALP with factored graphs.  
The approximated problem yields a low computational complexity. Under mild assumptions on the dependence structures and the separability of reward and control cost functions at each node, the resilient control policies can be designed in a distributed fashion. We can show that the system performance under the distributed control scheme is equivalent to the one achieved under the centralized one. We corroborate the established MDP model and analytical results with case studies of an interdependent power and subway network.
  
\textbf{Organization of the Paper}: The rest of the paper is organized as follows. Section \ref{model} establishes the MDP model of the interdependent infrastructures. Problem analysis is given in Section \ref{analysis}. The structural results including the optimal and distributed resilient control policies are obtained in Section \ref{newsetting}. The case study of a power and transit network is presented in Section \ref{case_study}, and Section \ref{conclusion} concludes the paper.


\section{SYSTEM MODEL}\label{model}
In this section, we introduce the network model to represent the complex interdependent SoS and the MDP model to produce resilient plannings of ICIs. The resilient plannings are both \textbf{proactive}, i.e., maintain high-risk nodes in advance of failures to reduce the future risks, and \textbf{cost-effective}, i.e., delay the repair of low-influence nodes due to the restrained or costly resources. 

\subsection{Network Model}
ICIs can be modeled by a multi-layer network as shown in Fig. \ref{multilayer}, where nodes represent heterogeneous components in multiple infrastructures such as buses in power grid, subway stations in transportation, and base stations in the communication network. The links connecting nodes in the network represent the dependencies either homogeneous, e.g., the failure of one subway station can disrupt the service of the next station, or heterogeneous, e.g., the power loss of the grid leads to the  shutdown of subway stations. Moreover, the inter-dependency implies the ping-pong effect. For example, the shutdown of subway stations can, in turn, slow down the power recovery because maintainers may take extra time to commute. The vicious spiral can lead to the cascading failure.
Without loss of generality, we use an aggregated network
$\mathcal{G}=(\mathcal{N},\mathcal{E})$ to capture the heterogeneous components in the multi-layer network, where $\mathcal{N}$ denotes a set of nodes and $\mathcal{E}$ is a set of links.


\subsection{MDP Model}
Each infrastructure node $i \in \mathcal{N}$ in the network $\mathcal{G}$ has a state $X_i$ representing its working status.  State $X_i$ admits values from a binary set $\{0,1\}$. Specifically, $X_i=1$ indicates that node $i$ works normally and $X_i=0$ means otherwise. In a network of $|\mathcal{N}|=n$ nodes, the system state can be denoted by $\mathbf{X}=[X_i]_{i \in \mathcal{N}} \in \mathcal{X}:=\{0,1\}^n$.
Due to the interdependencies between different nodes, the failure probability of one component depends not only on its own state but also on its connected ones.
Similarly, each infrastructure node $i\in\mathcal{N}$ can be controlled by taking action $A_i \in \{0,1\}$. Specifically, $A_i=1$ represents repair (if node is faulty) or maintenance (if normal) of node $i$, and $A_i=0$ indicates that no action is applied. The system action is a vector $\mathbf{A}=[A_i]_{i\in\mathcal{N}} \in \mathcal{A}:=\{0,1\}^n$ including all the actions of each node.

In order to design resilient mechanisms for the real-time operation of ICIs, we use an MDP model to capture the dynamics of the infrastructure states as well as the underlying failure and recovery processes of each node. Our objective is to determine a stationary policy $\pi: \mathcal{X}\mapsto \mathcal{A}$ that yields the optimal control of nodes to achieve the largest long-term benefit of ICIs. 

To achieve this goal, we further define the reward function $R: \mathcal{X}\times \mathcal{A} \mapsto \mathcal{R}$ and the objective function $V_\pi(\mathbf{x}): \mathcal{X}\mapsto \mathbb{R}$. The reward function measures the performance of the ICIs at one step by taking the number of working nodes into account. The objective function accumulates the obtained reward at each step over an infinite horizon: $V_{\pi}(\mathbf{x})=\mathbb{E}_{\pi}[\sum_{t=0}^{\infty}\gamma^{t}R\left(\mathbf{X}^{t},\pi(\mathbf{X}^{t})\right)|\mathbf{X}^{0}=\mathbf{x\mathrm{]}}$, where $\gamma<1$ is a positive discounted factor, $\mathbf{x}$ is an initial state, and  $\mathbf{X}^t$ is a random variable denoting the system state at stage $t$. The state transition probability follows the Markov property, i.e., the system state at the next time step only depends on the current system state and the adopted action. For example, $P(\mathbf{x}'|\mathbf{x},\mathbf{a})=\mathbb{P}\{\mathbf{X}^{t+1}=\mathbf{x}'| \mathbf{X}^t=\mathbf{x},\mathbf{A}^t=\mathbf{a} \}, \ \forall t$, denotes the transition probability from state $\mathbf{x} \in \mathcal{X}$ to $\mathbf{x}' \in \mathcal{X}$ under control $\mathbf{a} \in \mathcal{A}$. Note that transition probability $P$ captures the interdependencies between connected infrastructure nodes.
Further, the value function $V(\mathbf{x})=\max_\pi V_\pi(\mathbf{x}):=V_{\pi^*}(\mathbf{x})$ is the maximum achievable economic benefit of the ICIs starting from state $\mathbf{x}$ under the optimal policy $\pi^*$. 
A larger value of $V(\mathbf{x})$, $\forall \mathbf{x}$, indicates a better performance of ICIs, and the network is resilient to failures by using the control policy $\pi^*$.

\section{PROBLEM ANALYSIS}\label{analysis}
In this section, we introduce a sequence of solution techniques of the MDP problem
and tackle the curse of dimension by exploiting the sparse nature of the large-scale infrastructure networks. Approximation with the factor graph and the variable elimination technique leads to equivalent distributed policies. 

\subsection{Linear Programming}
One way to characterize the optimal policy of the MDP is via dynamic programming with the Bellman operator $(TV)(\mathbf{x}):=
\max_\pi R(\mathbf{x},\pi(\mathbf{x}))+\gamma\sum_{\mathbf{x}'\in\mathcal{X}}P(\mathbf{x}'|\mathbf{x},\pi(\mathbf{x}))V(\mathbf{x}'),\ \forall \mathbf{x}$.
Then, the value function $V(\mathbf{x})$ is the fixed point of $T$, i.e., $V(\mathbf{x})=(TV)(\mathbf{x}),  \forall \mathbf{x}$. 
Linear programming (LP) provides a convenient and efficient approach to solve the Bellman equation with value functions as LP variables, i.e., 
\begin{equation}
\begin{split}
\mathrm{(LP):}\ \min_{V(\mathbf{x})}:\quad & \sum_{\mathbf{x}\in\mathcal{X}}\alpha(\mathbf{x})V(\mathbf{x})\\
\mathrm{s.t.} \quad & V(\mathbf{x})\ge R(\mathbf{x},\mathbf{a})+\gamma\sum_{\mathbf{x}' \in \mathcal{X}}P(\mathbf{x}'|\mathbf{x},\mathbf{a})V(\mathbf{x}),\\
&\forall\mathbf{x}\in\mathcal{X},\ \forall \mathbf{a}\in \mathcal{A}.
\end{split}
\label{eq:exact LP}
\end{equation}
LP has the benefit to be easily extended to include both endogenous and exogenous constraints. For example, if we add an endogenous action constraint $\sum a_i=1$ which limits to single repair due to the limited resources at each stage, the LP formulation remains the same with only a reduced dimension of action space $\mathcal{A}'$ as in \shortcite{hln}. Similarly, we can directly add exogenous action constraints such as $\sum c_i(a_i)<C$ which gives a bound on the action cost due to limited budges at each step.

The disadvantage of \eqref{eq:exact LP} is that the curse of dimension appears in both state space $|\mathcal{X}|=2^n:=N$ and action space $|\mathcal{A}|=2^n:=N$ when solving the LP directly. Therefore, we use ALP and FMDP method to reduce the computation complexity \shortcite{guestrin2003efficient}.

First, we use the linear value function \shortcite{de2002linear} to approximate the value functions $V(\mathbf{x})=\sum_{i=1}^{k}w_{i}h_{i}(\mathbf{x})$, where $h_i$ and $w_i$ is the $i^{th}$ basis function and its weight, respectively. Hence the number of LP variables is reduced from $N=2^n$ to $k$ and we obtain the ALP with new variables $\{ w_1,w_2,...,w_k\}$.

\begin{equation}
\begin{split}
\mathrm{(ALP):}\ \min_{\mathbf{w}}\ & \sum_{\mathbf{x} \in \mathcal{X}}\alpha(\mathbf{x})\sum_{i \in \mathcal{N}}w_{i}h_{i}(\mathbf{x})\\
\mathrm{s.t.}\  \sum_{i \in \mathcal{N}}w_{i}&h_{i}(\mathbf{x})\ge R(\mathbf{x},\mathbf{a})+\gamma\sum_{\mathbf{x}'\in \mathcal{X}}P(\mathbf{x}'|\mathbf{x},\mathbf{a})\sum_{i \in \mathcal{N}}w_{i}h_{i}(\mathbf{x}'),\\
\forall\mathbf{x}\in&\mathcal{X},\ \forall \mathbf{a}\in \mathcal{A}.
\end{split}
\end{equation}
Our second step is to reduce the dimension of the transition probability matrix $P(\mathbf{x}'|\mathbf{x},\mathbf{a}) \in \mathcal{R}^{N\times N \times N}$. 
 The system transition probability $P(\mathbf{x'}|\mathbf{x},\mathbf{a})$ can be factored into a multiplication of local transition probabilities $\prod_{i=1}^n P(x_i|\mathbf{{x}}_{\bar{\Omega}_i})$ where  ${\bar{\Omega}_i}$ is the node set that can affect the local transition probability of node $i$. With the assumption that the network $\mathcal{G}$ is sparse, we have $|\bar{\Omega}_i|<n$.
Notice that the co-domain of each basis function $h_i(\mathbf{x})$ is determined by some subset of $\mathcal{X}$. The effective domain of $g_{i}(\mathbf{x},\mathbf{a}):=\sum_{\mathbf{x}'}P(\mathbf{x}'|\mathbf{x},\mathbf{a})h_{i}(\mathbf{x}')$ is also restricted to a set much smaller than $\mathcal{X}$ and a special case is discussed under assumption \ref{actionassumption}.
Define $\bar{g}_i(\mathbf{x},\mathbf{a}):=\gamma g_{i}(\mathbf{x},\mathbf{a})-h_{i}(\mathbf{x})$, 
the constraints $\sum_{i}w_{i} h_{i}(\mathbf{x})\ge R(\mathbf{x},\mathbf{a})+\gamma \sum_{i} w_{i}\ g_{i}(\mathbf{x},\mathbf{a}), \forall \mathbf{x},\mathbf{a}$ is rewritten in a more compact form $
0\geq \max_{\mathbf{x}, \mathbf{a}} \  R(\mathbf{x},\mathbf{a})+\sum_{i} w_{i} \ \bar{g}_{i}(\mathbf{x},\mathbf{a}),  \forall \mathbf{a} \in \mathcal{A}.$

The reward function $R(\mathbf{x},\mathbf{a})$ can also be factored into a sum of functions with the domain restrained to some subsets of $\mathcal{X}$. Then, we can use the variable elimination method to eliminate each element of $\mathbf{x}$ and $\mathbf{a}$ step by step \shortcite{hln}. 

\subsection{Approximate Optimal Policy}
The optimal centralized policy under the approximate form of LP is obtained by maximizing the objective function given the initial state $\mathbf{x}$, i.e., 
$
\mathbf{a}^*
\in \mathrm{arg}\max_{\mathbf{a} \in \mathcal{A}}
[R(\mathbf{x},\mathbf{a})+\gamma\sum_{\mathbf{x}' \in \mathcal{X}}P(\mathbf{x}'|\mathbf{x},\mathbf{a})V(\mathbf{x}')].
$
Similarly, an approximation is adopted to reduce the computation to $k+1$ multiplications and $k$ summations for any given pair of $(\mathbf{x},\mathbf{a})$, i.e., 
\begin{equation}
\begin{split}
\mathbf{a}^*& \in \mathrm{arg}\max_{\mathbf{a} \in \mathcal{A}}\left[R(\mathbf{x},\mathbf{a})+\gamma\sum_{i=1}^k w_i g_i(\mathbf{x},\mathbf{a})\right].
\end{split}
\label{eq:greedy}
\end{equation}
To solve \eqref{eq:greedy}, for every state $\mathbf{x}$, we need to search for action $\mathbf{a}$, which means that we have to compute (\ref{eq:greedy}) for $|\mathcal{X}|\times |\mathcal{A}|=2^{2n}$ times. However, by exploiting the structure of the problem, we can make assumptions in Section \ref{newsetting} to achieve a distributed computation of the optimal policy. 

\subsection{Main Structural Results}
\label{newsetting}
To study the interdependent infrastructure network consisting of $n$ nodes, we choose the basis function $h_i(\mathbf{x})$ to be the indicator function of node $i$'s state, i.e., $h_i(x_i,\mathbf{x}_{-i})=x_i, \forall \mathbf{x}_{-i} \in \{0,1\}^{n-1}$. Hence the number of basis functions is of the same size of the nodes, i.e.,  $k=n$. Weight parameter $w_i$ shows the importance of node $i$ in the measure of economic benefits. 

The advantage of choosing such basis functions is that the value function $V(\mathbf{x})$ at state $\mathbf{x}=[x_i]_{i \in \mathcal{N}}$ is approximated by the summation of the weight $w_i$ of all working node $x_i=1$ at state $\mathbf{x}$. With a larger number of critical nodes functioning, the system achieves a higher reward in the long run.
This choice of basis functions is particularly suitable for sparse networks  because they primarily capture local effects. We can add new basis functions such as $h_i(x_i,x_j,\mathbf{x}_{\mathcal{N}\backslash\{i,j\}})= x_i \cdot x_j, \forall j \in \Omega_i$, which is necessary when non-local effects play a major role.
We make the following reasonable assumptions to achieve a distributed repair policy.
\begin{assumption} 
The local transition probability of node $i$ is not affected by the control value of other nodes, i.e., 
$P(x_i|\mathbf{{x}}_{\bar{\Omega}_i},\mathbf{a})=P(x_i|\mathbf{{x}}_{\bar{\Omega}_i},a_i)$.
\label{actionassumption}
\end{assumption}
Under the above assumption, we obtain the following equations:
\begin{eqnarray}
\nonumber g_{i}(\mathbf{x},\mathbf{a}) &=&\sum_{\mathbf{x}' \in \mathcal{X}}P(\mathbf{x}'|\mathbf{x},\mathbf{a})h_{i}(x_{i}')
=\sum_{i \in \mathcal{N}} \prod_{j \in \mathcal{N}} P(x_{j}'|\mathbf{{x}}_{\bar{\Omega}_{j}},\mathbf{a})h_{i}(x_{i}')\\
\nonumber &=&\sum_{x_{i}'} P(x_{i}'|x_{i},\mathbf{x}_{\Omega_{i}},\mathbf{a})h_{i}(x_{i}')
=P(X_{i}^{'}=1|\mathbf{{x}}_{\bar{\Omega}_{i}},a_i).
\end{eqnarray}

\begin{assumption}
We can separate and factor the reward function $R(\mathbf{x},\mathbf{a})$ into the net reward $\sum_{i=1}^n r_i(\mathbf{{x}}_{\bar{\Omega}_i})$ and the action cost $-\sum_{i=1}^n c_i(a_i)$, where $r_i(\mathbf{{x}}_{\bar{\Omega}_i})$ and $c_i(a_i)$ are the local reward and the cost function of node $i$, respectively.
\label{separatereward}
\end{assumption}
\begin{theorem}
\label{distributed policy}
 A distributed policy $a_i \in \mathrm{arg}\max_{a_i} \gamma w_i g_i(\mathbf{{x}}_{\bar{\Omega}_{i}},a_i)-c_i(a_i)$ achieves the same reward obtained by the centralized optimal policy using \eqref{eq:greedy} based on assumption \ref{actionassumption} and \ref{separatereward}.
\end{theorem}
Theorem. \ref{distributed policy} holds as we can represent the RHS of Eq. \eqref{eq:greedy} in a factored form, i.e., $$ \max_{a_1,...,a_n}[R(\mathbf{x},\mathbf{a})+\gamma\sum_i w_i g_i(\mathbf{x},\mathbf{a})]=\sum_{i=1}^k r_i(\mathbf{{x}}_{\bar{\Omega}_i})+
\sum_{i=1}^n \max_{a_i} \gamma w_i g_i(\mathbf{{x}}_{\bar{\Omega}_{i}},a_i)-c_i(a_i).$$ 
The distributed policy only takes into account the states of node $j \in \bar{\Omega}_i$ and reduces the computation complexity from $O(|\mathcal{X}|\times |\mathcal{A}|)$ to $n \cdot 2^{\Omega_i}$.
In many applications, the policy designer can only control a limited number of nodes. Thm. \ref{distributed policy} still holds as the sparse control condition can be viewed as a special case.

Our next step is to find a pattern to quickly identify each node's optimal action, i.e, repair or not. Thm. \ref{pattern} achieves the above objective under an arbitrarily given set of controllable nodes without explicitly computing $a_i \in \mathrm{arg}\max_{a_i} \gamma w_i g_i(\mathbf{{x}}_{\bar{\Omega}_{i}},a_i)-c_i(a_i)$.
With assumption \ref{nature}, we can find the repair policy thresholds as summarized in Thm. \ref{pattern}, which means that we switch to the opposite action if the condition is not satisfied. Let $\bar{\Omega}_i=\{ i,\Omega_i\}$ and $\mathbf{{x}}_{\bar{\Omega}_i}=[x_i,\mathbf{x}_{\Omega_i}]$. Besides, we define a shorthand notation $P_ {\mathbf{x}_{\Omega_i}=j}:=P(X_i^{'}=1 | \mathbf{X}_{\Omega_i}=j,A_i=0)$. 
\begin{assumption}
We further assume that a repair is always effective $P(X_i^{'}=1 | \mathbf{x},a_i=1)=1,  \forall i, \forall \mathbf{x}$ and the node stays faulty without a repair $P(X_i^{'}=1 | x_i, \mathbf{x}_{\Omega_i},a_i=0)=x_i \cdot P_ {\mathbf{x}_{\Omega_i}},  \forall i, \forall \mathbf{x}$. In addition, the cascading failure means that $P_{\mathbf{{x}}_{\Omega_i}=\mathbf{1}}>P_{\mathbf{{x}}_{\Omega_i} \neq \mathbf{1}}$, where $\mathbf{1}$ represents a vector with all $n$ elements as $1$.
\label{nature}
\end{assumption}
\begin{theorem}
With assumptions \ref{actionassumption}, \ref{separatereward} and \ref{nature}, we only repair node $i$ when it is faulty and its action cost is relatively low, i.e., when $c_i$ satisfies $c_i(1)-c_i(0)<w_i$.
If $0<c_i(1)-c_i(0)<w_i(1-P_{\mathbf{{x}}_{\bar{\Omega}_i}=\mathbf{1}})$, we repair node $i$ even when nodes in ${\bar{\Omega}_i}$ are all working normally.
We choose to repair node $i$, i.e., $A_i=1$, when $i$ is working $X_i=1$ yet other nodes in $\Omega_i$ have states $\mathbf{x}_{\Omega_i} \neq \mathbf{1}$ under the condition $w_i(1-P_{\mathbf{{x}}_{\bar{\Omega}_i}=\mathbf{1}})<c_i(1)-c_i(0)<w_i(1-P_{\mathbf{{x}}_{\Omega_i} \neq \mathbf{1}})$.
\label{pattern}
\end{theorem}
To show Thm. \ref{pattern}, we first look at the case when only one given node is controllable.
Since action $a_i$ is binary, we can obtain the optimal policy by comparing which action leads to a larger value.
We generalize the one-node case to the case of multiple nodes because the Thm. \ref{distributed policy} guarantees that each node's policy can be decided independently of others' actions.

\section{CASE STUDY}
\label{case_study}

\begin{figure}
\includegraphics[width=0.6\linewidth]{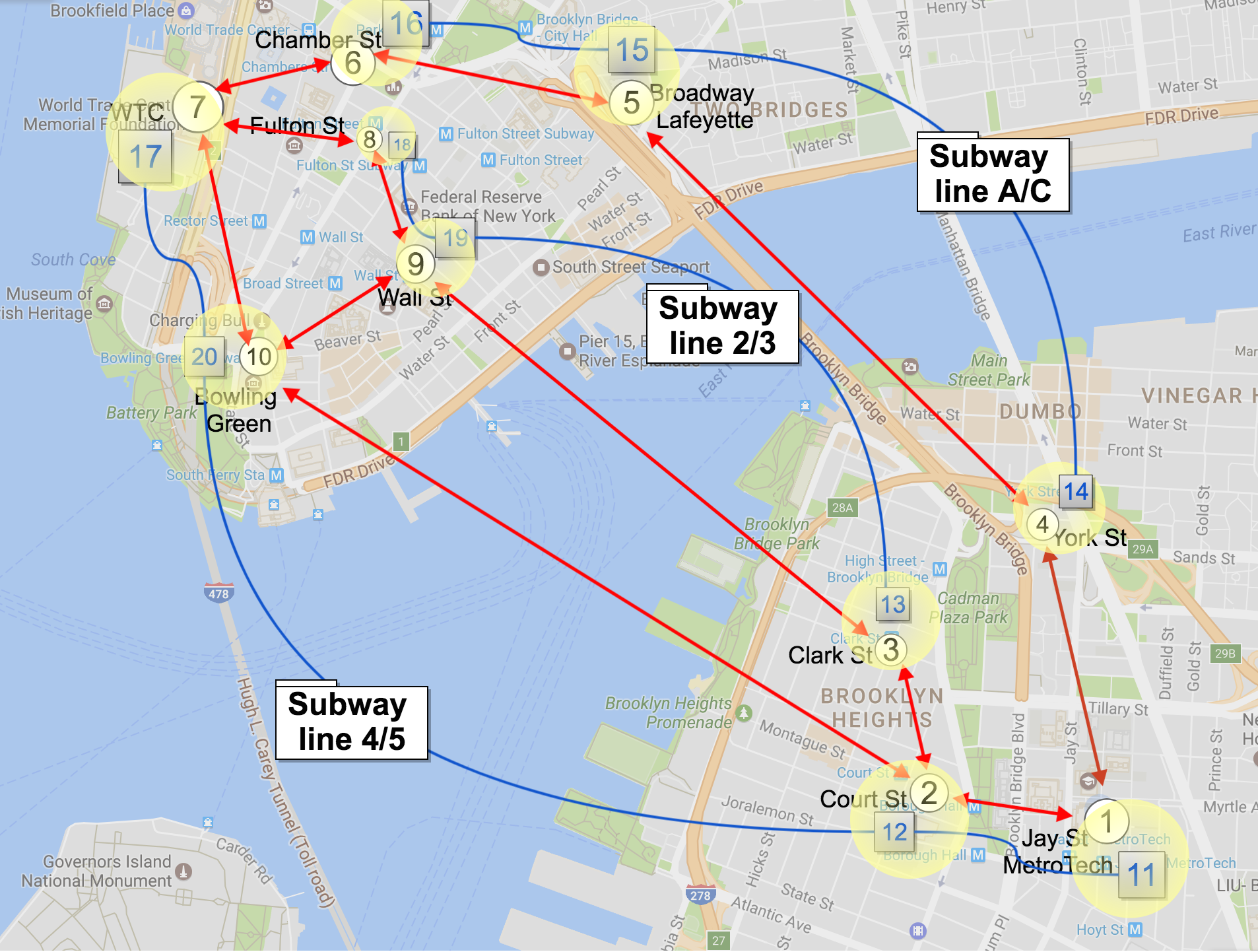}
\centering
\caption{Network representation of the suggested interdependent microgirds and subway system at the lower Manhattan and downtown Brooklyn area. Red and blue links represent power and subway lines, respectively. Power nodes are indexed from 1 to 10 while the coupled subway stations are indexed from 11 to 20. The size of node $i$ is proportional to its reward $r_i(\mathbf{x}_{{\bar{\Omega}_i}})$, indicated by the people flow density data from  \href{http://web.mta.info/nyct/facts/ridership/ridership_sub_annual.htm}{MTA}. Power and subway nodes of the same place inside one yellow circle share the same local reward. }
\label{bidirection}
\end{figure}

In this section, we use a case study motivated by Hurricane Sandy, which struck New Jersey on Oct. 29, 2012, resulting in an estimated \$50 billion economic loss and at least 147 direct deaths \shortcite{blakeii}. It wreaked havoc on critical infrastructures such as power grid, transportation and communication systems. In this case study, we focus on the impact of the power and subway failures in the lower Manhattan and downtown Brooklyn as shown in Fig. \ref{bidirection}\footnote{Background image source: Google map}.

\subsection{Power and Subway}
The most direct impact of Sandy has been on the power grid. Much of Manhattan south of 39th Street has suffered a massive power cut because of the floods and the high wind. A \href{http://www.huffingtonpost.com/2012/10/30/coned-explosion-hurricane-sandy-video_n_2044097.html}{blast of one Con Edison substation} on 14th Street at 9 p.m. on Oct. 29 has intensified the power outage. 
Besides, ConEd took parts of its grid offline due to the rising water. Load shedding arose when the demand strained the capacity during the storm. The massive damages made ConEd  unable to restore most of the power outages in lower Manhattan until 2 November.
The whole power system has taken weeks to recover completely, and nearly  8.5  million people have suffered from power outages. 

\begin{figure}
\includegraphics[width=0.6 \linewidth]{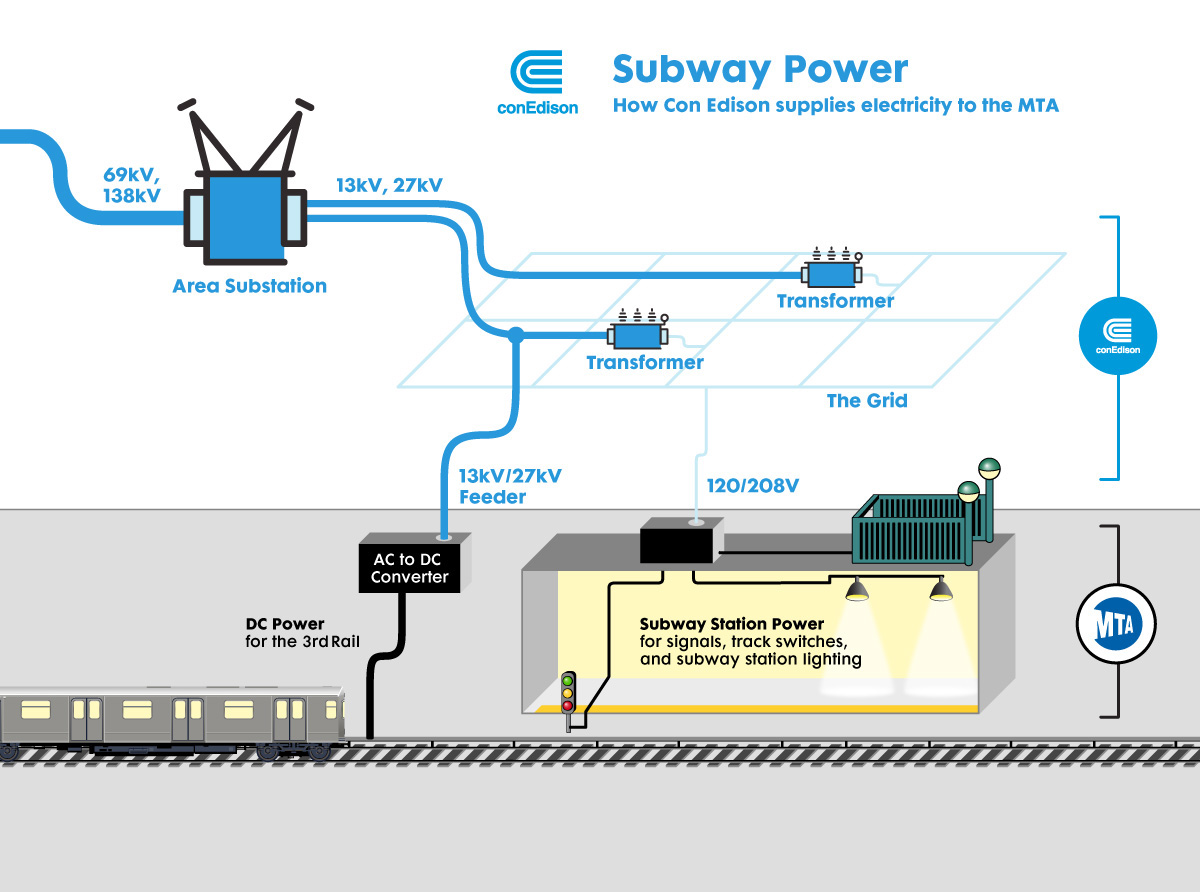}
\centering
\caption{After the high-voltage transmission of 69/138kV,  area substations transform the power voltage to 13/27kV, convert it to direct power current and feed it for the 600-volt third rail of the track to propel the train. Besides, transformers in the distribution grid further lower the voltage to 120/208V to power signals, lights, and track switches in subway stations.}
\label{ConMTA}
\end{figure}

As shown in Fig. \ref{ConMTA}\footnote{Source: https://www.coned.com/en/about-us/media-center/multimedia-library-or-gallery.}, the normal operation of subway system relies heavily on a stable power supply. 
Thus, the subway system has been significantly affected as well. 
The MTA shuts down the entire subway system from 7 p.m. on Oct. 28  to move trains away from lowland and vulnerable area. 
Floodwaters brought by the storm surge of Sandy began to enter the subway tunnels and stations after 9 p.m. on 29 October. 
All seven East River subway tunnels connecting Manhattan, Brooklyn, and Queens have suffered from floods a day later. Besides, Hurricane Sandy wiped out tracks on the A train in the Rockaways and severely damaged new South Ferry station. 

Since after flooding, water can short-circuit electrical devices including signals, switches, and the electrified third rail, the subway has to remain closed before dewatering. The pump system of the normal drainage does not function during Sandy because of the power cut. Thus, it takes a long time to pump out water with limited backup power generators. After seawater has been pumped out, MTA needs extra effort to clean the salt deposits and debris, inspect, test and repair the electrical components to guarantee the safety of train operations. 
Unfortunately, not only the resilience of electric power has a significant impact on the recovery of the subway system, the repairing process of electric power systems is strongly influenced by the subway as well because the subway shutdown creates difficulty for ConEd crews to reach faulty components. 

\subsection{Network Construction}
Lessons learned from Sandy has indicated that the lack of sufficient backup power decreases the recovery efficiency due to the failure of conventional pumping systems. Hence, backup generators (e.g., diesel generators) or local energy storage devices should be prepared in case of power failures. For instance, Verizon's Garden City central office powered by seven fuel cells has saved it from the loss of power \shortcite{kwasinski2013lessons}. Thus, 
Microgrids can be adopted to supply power to local users when the centralized power distribution network breaks down as what has happened during Sandy. 
In this case study, we construct a 20-node interdependent network illustrated in Fig. \ref{bidirection}, where each microgrid node $1$-$10$ represents a local backup generator located at the corresponding subway station $11$-$20$. Transmission lines in red connect the microgrid so that loads of each node can change due to its working status. The blue lines construct three simplified subway lines. Let $X_i=1$ be the state of normal operation while $X_i=0$ indicates the failure of node $i$.

Backups can fail and 
we set up the failure model of the microgrids based on our previous work \shortcite{hln}. The failure of microgrids at subway stations can lead to the meltdown of the public transit system. For example, if subway station $17$ is out of service due to the power outage of node $7$, the entire line $4/5$ has to suspend as trains cannot get through.  On the other hand, the subway system affects the microgrids through reward $r_i(\mathbf{x}_{{\bar{\Omega}_i}})$ of node $i$. 
We adopt the proposed framework to illustrate how we achieve the resilience of the microgrids and the subway lines under our optimal policy. For each node $i$, $A_i=1$ means to repair or maintain the node based on the state of the node. On the other hand, $A_i=0$ means that no action is taken.
 
The coupling of the power and subway system exists as an illustration of the interdependent infrastructures. The failure of microgrids at subway stations can lead to the meltdown of the public transit system. The subway system affects the microgrids through reward $r_i(\mathbf{x}_{{\bar{\Omega}_i}})$ to node $i$.

\subsection{Computation Result}

\begin{table}[t]
\centering
\caption{Results of ALP with all nodes controllable.}
\label{Computation Results}
\begin{tabular}{|c|c|c|c|c|}
\hline
\multicolumn{3}{|c|}{Node Index and Name} & Local Reward $r_i(\mathbf{x}_{{\bar{\Omega}_i}})$ & Weight $w_i$\\ \hline
1 & Jay St. & 11 & 15.3215 & 18.5749 \\ \hline
2 & Court St. & 12 & 7.1197 & 8.7800 \\ \hline
3 & Clark St. & 13 & 2.4738 & 3.8601 \\ \hline
4 & High St. & 14 & 2.9452 & 3.9617 \\ \hline
5 & Broadway & 15 & 15.7205 & 17.5535 \\ \hline
6 & Chamber St. & 16 & 14.1986 & 14.2076 \\ \hline
7 & WTC & 17 & 20.4828 & 25.2162 \\ \hline
8 & Fulton St. & 18 & 1.7065 & 2.6374 \\ \hline
9 & Wall St. & 19 & 9.3785 & 11.2510 \\ \hline
10 & Bowling Gr & 20 & 10.6528 & 12.8872 \\ \hline
\end{tabular}
\end{table}

Table \ref{Computation Results} shows that the weight of each node obtained by ALP follows the same ordering as the local reward except for node 1 and 5. This result arises from the fact that the topology of the network has less influence on the weight than the local reward.
The centralized and the distributed search methods yield the same policy as indicated by Thm. \ref{distributed policy} and the optimal policy satisfies the pattern stated in Thm. \ref{pattern}. The optimal policy is not to just myopically repair defective nodes but also maintain some healthy nodes to achieve a long-term system-wide performance. 
For instance, we observe that the policy of the most critical node (WTC) is not repaired only when node $6$ and $7$ are both working as shown in Fig. \ref{recoverymap}. We choose to maintain node $7$ when $X_6=0, X_7=1$ to prevent node $7$'s outage in later stages. Other neighboring nodes $6$ and $8$ are not influential on the policy of node $7$ because they have relatively small weights as shown in Table \ref{Computation Results}.
As for the computation complexity, it takes the centralized search method about $13843.7$ seconds to solve the case after weights $w$ are computed while our distributed search process requires only $2.1$ seconds. 

\subsection{Simulation Result}

\begin{figure}
\subfigure{
\includegraphics[width=0.49\columnwidth]{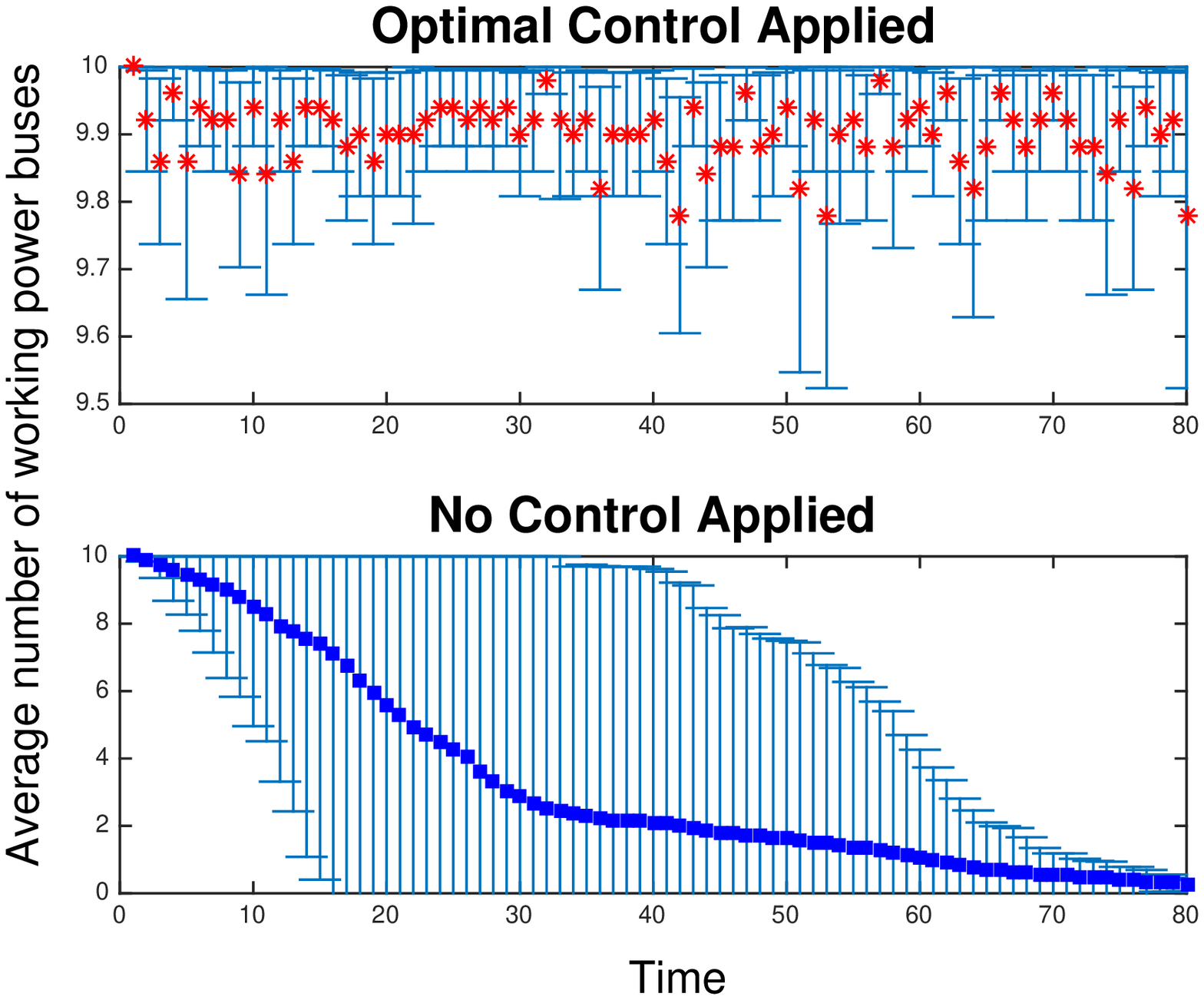}
}
\subfigure{
\includegraphics[width=0.49\columnwidth]{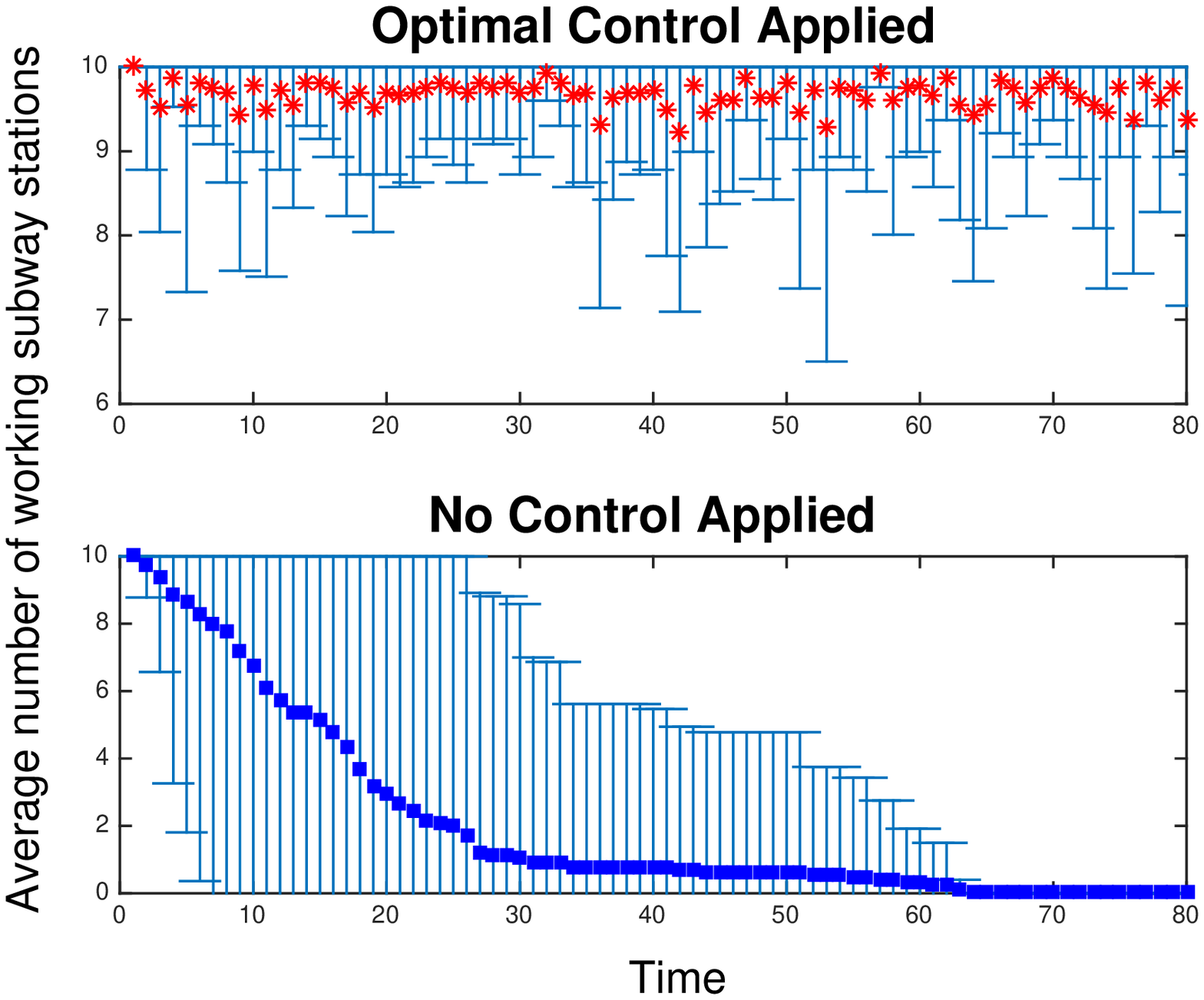}
}
\caption{The resilience of the microgrid and subway system with and without the optimal policy. The dot indicates the mean value while the error bar shows the variance of the sample trajectory. Our optimal policy largely reduces the variance and thus makes the system more stable.}
\label{systemresilience}
\end{figure}

We choose $\mathbf{X}=\mathbf{1}$ as the initial state and show the system resilience under our optimal policy in Fig. \ref{systemresilience}. 
Averaged under 50 iterations, the data of the $y$-axis shows the average number of working nodes and also measures the resilience of the system. Without controls, although the prior failure probability is set to a small number $0.01$, cascading failure still occurs, and the survival number of nodes quickly decreases to zero. On the other hand, our optimal policy repairs the proper nodes so that the system quickly recovers and stays at the healthy state $\mathbf{X}=\mathbf{1}$. The variance of the subway system is larger than that of the microgrid because one subway station out of service makes the entire line break down.


\begin{figure*}[h]
\begin{multicols}{3}
\includegraphics[width=\linewidth]{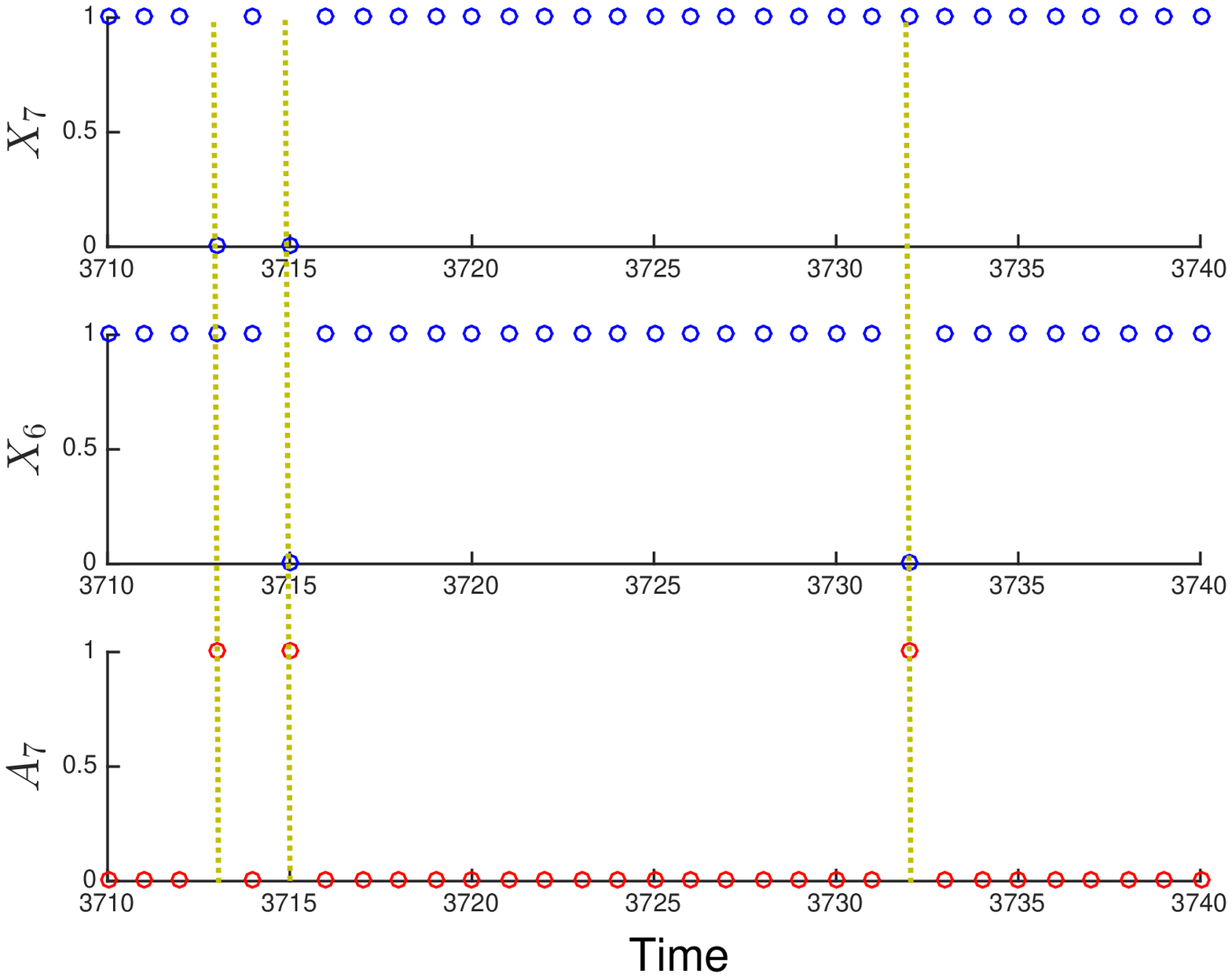}\par\caption{The blue circles at the top two subfigures show the state transition of node $7$ and $6$ respectively. The subplot at the bottom shows the corresponding action of node $7$.}\label{recoverymap}

    \includegraphics[width=\linewidth]{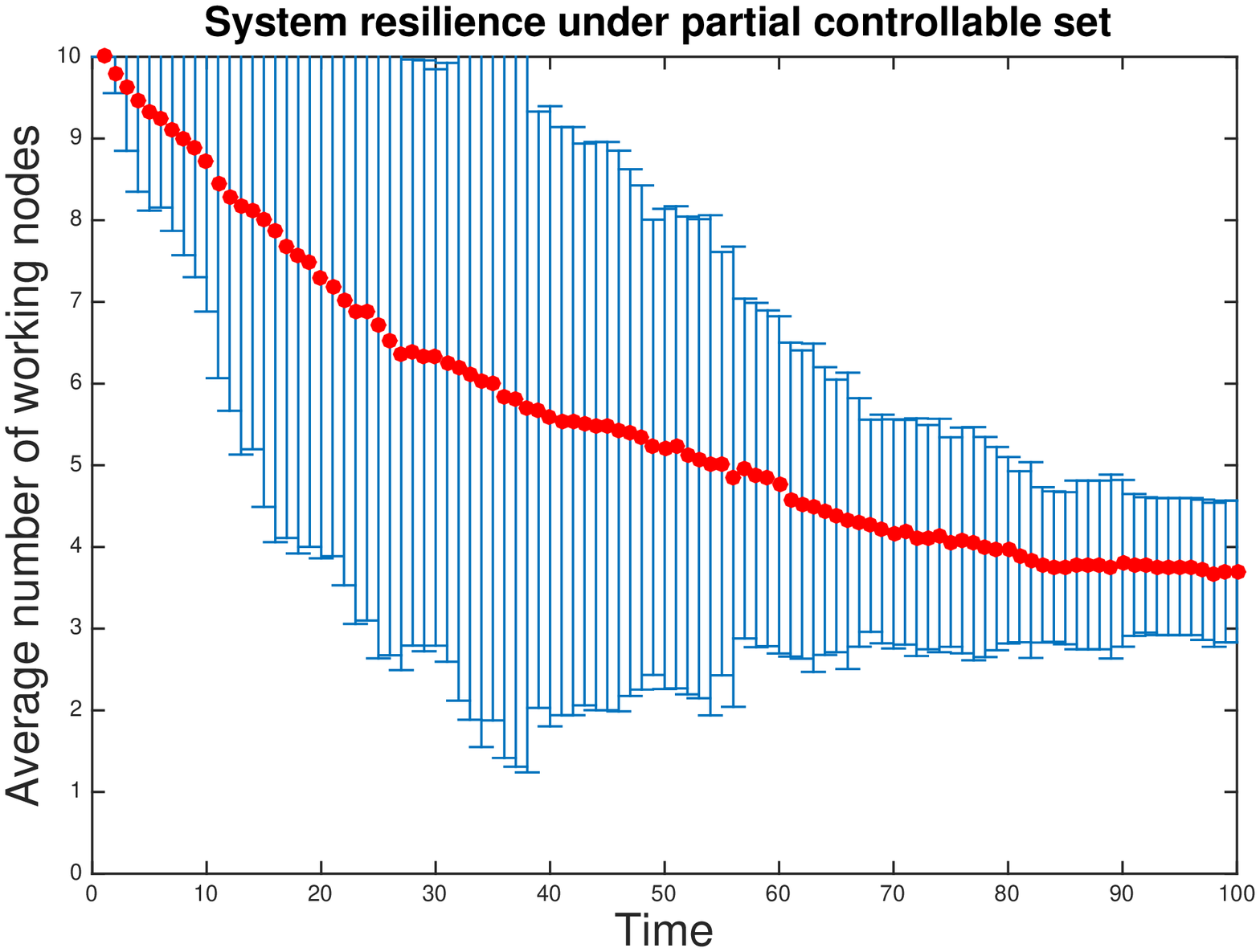}\par\caption{Variations of the average number of working nodes when only node $7$ (WTC) and node $8$ (Fulton St.) are controllable. The red dots indicate an average number of the working node with error bars shown in blue.}\label{givencontrol}
    \includegraphics[width=0.9\linewidth]{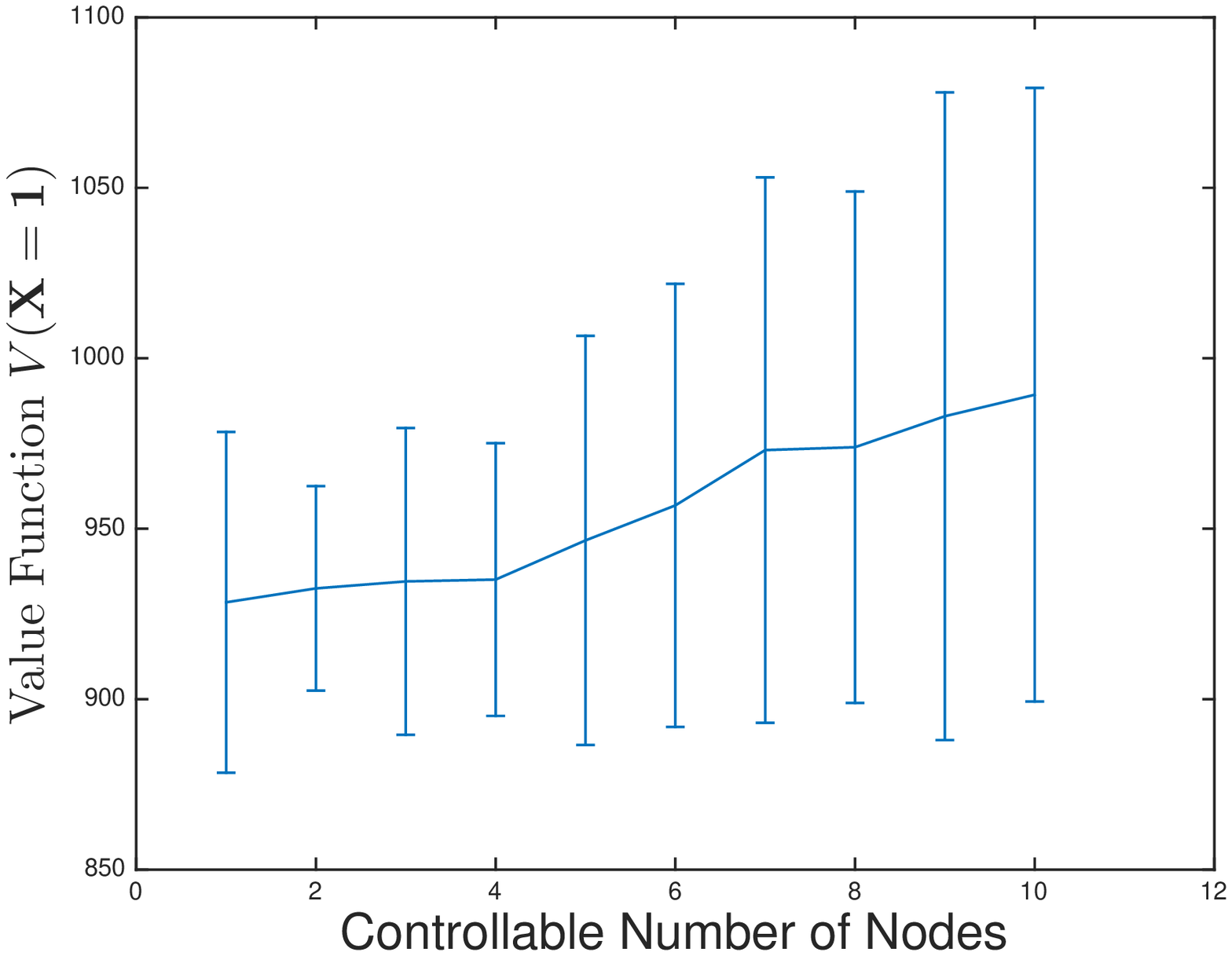}\par\caption{System reward grows with the increase of the number of controllable nodes. The error bar shows the upper and the lower bounds adopting different combinations of controllable nodes.}
    \label{increasevalue}
\end{multicols}
\end{figure*}

We compare three other policies. Policy $1$ is a straightforward strategy, i.e., to repair every faulty node and leave alone working nodes. Policy $2$ represents a random strategy that chooses to restore each outage node with probability $4/5$ and repair each non-outage node with probability $1/5$. Taking no action is applied as policy $3$.
Our simulation results show that policy $1,2,3$ achieve the value of $979.81$,$947.17$ and $796.87$ with initial state $\mathbf{X}=\mathbf{1}$, while our optimal policy achieves a higher value of $985.44$. 
The optimal value obtained by ALP is $989.2963$, which is close to the `actual value' obtained from the simulation. Note that it is computationally expensive to directly solve the exact LP. We can approximate the exact value by simulating over a long horizon and averaging over a sufficient large number of samples.

Finally, we study the case when only part of nodes is controllable. In Fig. \ref{givencontrol},  we maintain the resilience of some nodes with a sparse manageable set. For example, optimal control of nodes $6$ and $8$ keeps on average four nodes healthily working. More controllable nodes lead to a larger value function as shown in Fig. \ref{increasevalue} and the growth is approximately linear with the size of controllable nodes.

\subsection{Large-scale Demo }
We illustrate the cascading failure of a two-layer network\footnote{https://drive.google.com/drive/u/0/folders/0B6-Q8-SnvO6lYmlxX2FuN3ZuV1U} with 100 nodes representing mixed components. 
The failure probability of each node is proportional to the number of its connected faulty nodes and is illustrated via the color bar. A lighter color corresponds to a higher failure probability, and the  white color means the node has failed. If a node is repaired or maintained, then it will not fail in a few stages afterward and is shown in black. We introduce both edge-based and node-based resilience metrics, i.e., connectivity and the reward of the working nodes, respectively. They are normalized to 100 with their maximum values. 
As we can see from the video demo, the system starts with a two-node failure and propagates to the neighboring nodes quickly without effective controls. 

On the contrary, the dynamic recovery demo shows how the network can recover quickly and wholly from the catastrophic event where almost all nodes fail. Also, the resilient planning manages to maintain a stable, healthy state of the entire network in the long run after the recovery. We assume that the limited resources can only support one node recovery at each stage to obtain a clear visualization. 
Our optimal control policy achieves a tradeoff between a high-level resilience and the cost it takes.

\section{CONCLUSION}
\label{conclusion}
In this paper, we have formulated a factored MDP model for large-scale ICIs and significantly reduced the complexity of computing optimal policies. A distributed optimal control policy is designed to enhance the resilience and security of ICIs. 
The proposed framework has been applied to a case study motivated by Hurricane Sandy. We have shown that our policy manages to harden the security by reducing the failure probability and achieve resilience by acting on proper nodes. Moreover, the cost achieved by the found policies is the lowest. The framework developed in this work would be useful to mitigate large-scale infrastructure networks through optimal dynamic resilience planning. Even with a small subset of hardened nodes through planning, we have observed a sizable loss reduction for extreme events. We have used microgrids and subway systems in lower Manhattan and downtown Brooklyn as a case study to illustrate the significant resilience improvement in event of Hurricane Sandy. Our framework can be applied to develop resilience policies for other interdependent infrastructures including water distribution systems, food systems, data centers, and communication infrastructures. The future work would aim at understanding implementation and resource constraints on the policies and developing a large-scale solver that would provide usable infrastructure solutions.

\section*{ACKNOWLEDGMENTS}
This research is partially supported by a DHS grant through Critical Infrastructure Resilience Institute (CIRI) and grants SES-1541164, and ECCS-1550000 from National Science Foundation (NSF).

\bibliographystyle{wsc}
\bibliography{sig_ref}

\section*{AUTHOR BIOGRAPHIES}

\noindent {\bf Linan Huang} received the B.Eng. degree in Electrical Engineering from Beijing Institute of Technology, China, in 2016. He is currently a second-year Ph.D. candidate in the Laboratory for Agile and Resilient Complex Systems, Tandon School of Engineering, New York University, NY, USA. 
His research interests include optimal dynamic decision making of the multi-agent system,  cyber-physical security, and resilient infrastructure networks. His e-mail address is \email{lh2328@nyu.edu}. \\

\noindent {\bf Juntao Chen} received the B.Eng. degree in Electrical Engineering and Automation with honor from Central South University, Changsha, China, in 2014. He is currently pursuing the Ph.D. degree in the Laboratory for Agile and Resilient Complex Systems, Tandon School of Engineering, New York University, NY, USA. 
His research interests include game theory, cyber-physical systems, interdependent complex networks, Internet of things, and mechanism design. His e-mail address is \email{jc6412@nyu.edu}. \\

\noindent {\bf Quanyan Zhu} is an assistant professor in the Department of Electrical and Computer Engineering at New York University. He received the B. Eng. in Honors Electrical Engineering with distinction from McGill University in 2006, the M.A.Sc. from University of Toronto in 2008, and
the Ph.D. from the University of Illinois at Urbana Champaign (UIUC) in 2013. From 2013-2014, he was a postdoctoral research associate at the Department of Electrical Engineering, Princeton University. 
He is a recipient of many awards including NSERC Canada Graduate Scholarship (CGS), Mavis Future Faculty Fellowships, and NSERC Postdoctoral Fellowship (PDF). He spearheaded and chaired INFOCOM Workshop on Communications and Control on Smart Energy Systems (CCSES), Midwest Workshop on Control and Game Theory (WCGT), and 7th Game and Decision Theory for Cyber Security (GameSec). His current research interests include resilient and secure interdependent critical
infrastructures, energy systems, cyber-physical systems, and smart cities. His e-mail address is \email{qz494@nyu.edu}.

\end{document}